# Artificial Intelligence-Enabled Optimization of Battery-Grade Lithium Carbonate Production

S. Shayan Mousavi Masouleh [1, 2], Corey A. Sanz [3], Ryan P. Jansonius [3], Samuel Shi [4], Maria J. Gendron Romero [4], Jason E. Hein [3], Jason Hattrick-Simpers [1, *]

[1] Canmet MATERIALS, Natural Resources Canada, 183 Longwood Rd S, Hamilton, ON, Canada

[2] Department of Materials Science and Engineering, McMaster University, 1280 Main St W, Hamilton, ON, Canada

[3] Telescope Innovations, 301-2386 E Mall, Vancouver, BC, Canada

[4] Department of Materials Science and Engineering, University of Toronto, 184 College St, Toronto, ON, Canada

**Corresponding Author**

* jason.hattrick-simpers@nrcan-rncan.gc.ca

**ABSTRACT**

By 2035, the need for battery-grade lithium is expected to quadruple. About half of this lithium is currently sourced from brines and must be converted from a chloride into lithium carbonate ($Li_2CO_3$) through a process called softening. Conventional softening methods using sodium or potassium salts contribute to carbon emissions during reagent mining and battery manufacturing, exacerbating global warming. This study introduces an alternative approach using carbon dioxide ($CO_2(g)$) as the carbonating reagent in the lithium softening process, offering a carbon capture solution. We employed an active learning-driven high-throughput method to rapidly capture $CO_{2(g)}$ and convert it to lithium carbonate. The model was simplified by focusing on the elemental concentrations of C, Li, and N for practical measurement and tracking, avoiding the complexities of ion speciation equilibria. This approach led to an optimized lithium carbonate process that capitalizes on $CO_2(g)$ capture and improves the battery metal supply chain's carbon efficiency.

**TOC GRAPHICS**

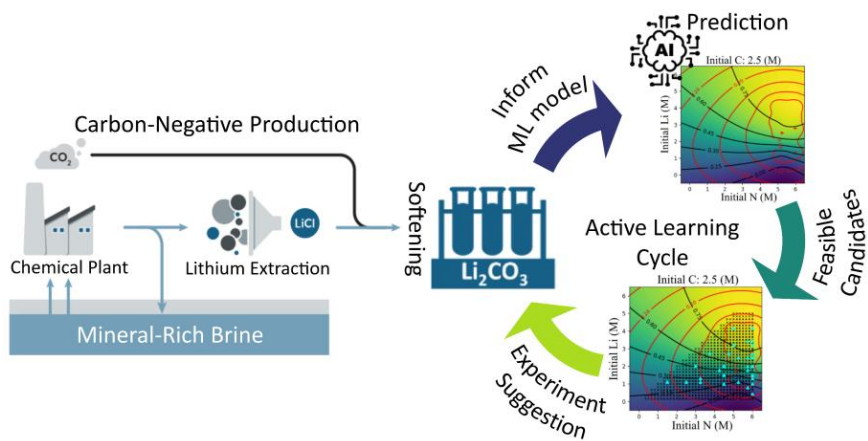

Lithium carbonate is a critical precursor for the production of lithium-ion batteries which range from use in portable electronics to electric vehicles. In fact, battery applications account for over 80% of all lithium produced globally and demand a high purity level, with raw lithium carbonate ($Li_2CO_3$) requiring a purity above 99% [1,2]. Typically, lithium is sourced from brines or clays in forms such as lithium chloride or sulfate which are then converted to lithium carbonate through a metathesis reaction. In this reaction, the chloride or sulphate anion is replaced by carbonate, precipitating the insoluble lithium carbonate product (see Equation 1) [1,3–5]. This process, called softening, must be performed such that impurities in the brine are excluded from the product during crystallization, and the crystal size is sufficiently large to facilitate filtration and isolation [1,3,6].

Conventional Method:   $2LiCl + Na_2CO_3 \rightleftharpoons 2NaCl + Li_2CO_3$      (Eq. 1)

Current Study:  $2LiCl + 2NH_4OH + CO_{2(g)} \rightleftharpoons Li_2CO_3 + 2NH_4Cl + H_2O$     (Eq. 2)

The conventional approach to softening usually uses mined sodium carbonate ($Na_2CO_3$) as the carbonate source, producing soluble sodium chloride (NaCl), or sodium sulfate ($Na_2SO_4$) as the reaction byproduct [3,4]. The presence of these sodium salts in the crystallization matrix can lead to inclusion within the $Li_2CO_3$ crystals, resulting in a lower purity of the final product [6,7]. Moreover, the use of sodium carbonate salt in battery manufacturing leads to exposure of carbonate ions to acids, releasing carbon dioxide gas ($CO_2(g)$) into the atmosphere. Additionally, the mining and transporting of sodium (or potassium) carbonate contributes to the overall carbon footprint of this process [7].

In contrast, converting onsite-produced $CO_{2(g)}$ into a carbonate source for softening can significantly diminish the carbon footprint of the process, provide an avenue to remove sodium from the solution and potentially improve the purity of the final product. One approach is to react $CO_{2(g)}$ with an aqueous solution of ammonium hydroxide ($NH_4OH$) to produce ammonium carbonate in-situ (($NH_4)_2CO_3$) [8,9]; solutions of lithium chloride (LiCl) can then be treated with this ammonium carbonate to produce lithium carbonate (the net equilibrium can be seen in Equation 2). However, softening lithium brines with ammonium carbonate involves complex equilibria governing $CO_{2(g)}$ dissolution, $Li_2CO_3$ precipitation, and the reaction of carbonic acid (a weak acid) with ammonium hydroxide (a weak base) [8,10,11]. The reaction produces ammonium as a byproduct, which is a weak acid that can protonate carbonate, hindering $Li_2CO_3$ precipitation by converting dissolved carbonate ions into bicarbonate [10]. A system of multiple inter-related equilibria describes this chemical system (**Table 1**).

**Table 1.** Detailed equilibrium reactions in lithium brine softening using ammonium hydroxide and carbon dioxide gas.

| Vapor-Liquid | | Solid Formation | |
|---|---|---|---|
| $CO_{2(g)} \rightleftharpoons CO_{2(aq)}$ | (Eq. 3) | $NH_4^+ + HCO_3^- \rightleftharpoons NH_4HCO_{3(s)}$ | (Eq. 11) |
| $NH_{3(g)} \rightleftharpoons NH_{3(aq)}$ | (Eq. 4) | $NH_4^+ + NH_2COO^- \rightleftharpoons NH_2COONH_{4(s)}$ | (Eq. 12) |
| $H_2O_{(g)} \rightleftharpoons H_2O$ | (Eq. 5) | $2NH_4^+ + CO_3^{2-} + HCO_3^- \rightleftharpoons (NH_4)_2CO_3 \cdot H_2O_{(s)}$ | (Eq. 13) |
| **Liquid Speciation** | | $4NH_4^+ + CO_3^{2-} + 2HCO_3^- \rightleftharpoons (NH_4)_2CO_3 \cdot 2NH_4HCO_{3(s)}$ | (Eq. 14) |
| $NH_{3(aq)} + H_2O \rightleftharpoons NH_4^+ + OH^-$ | (Eq. 6) | **Carbonate Formation** | |
| $CO_{2(aq)} + H_2O \rightleftharpoons H^+ + HCO_3^-$ | (Eq. 7) | $H_2CO_3 \rightleftharpoons H^+ + HCO_3^-$ | (Eq. 15) |
| $HCO_3^- \rightleftharpoons H^+ + CO_3^{2-}$ | (Eq. 8) | | |
| $H_2O \rightleftharpoons H^+ + OH^-$ | (Eq. 9) | $2Li^+ + CO_3^{2-} \rightleftharpoons Li_2CO_{3(s)}$ | (Eq. 16) |
| $NH_{3(aq)} + HCO_3^- \rightleftharpoons NH_2COO^- + H_2O$ | (Eq. 10) | | |

The goal of this work is to identify a set of reaction conditions that maximizes lithium carbonate formation (Eq.16). Modeling this complex system of equilibria through traditional thermodynamic approaches is challenged by the multitude of independent chemical species, the numerous inter-related reactions, and the possible importance of kinetics [7]. Experimental approaches to optimize this system were similarly challenged by the complexity of the search space and its potential nonlinearities. Among the few examples of $Li_2CO_3$ production with $CO_2(g)$, M. Tian et al. observed that increasing ammonium hydroxide concentration from 200 g/L to 400 g/L raised the yield from 43.0% to 49.6% [12]. The highest reported yield in similar systems (using LiOH) is about 73% with 2 mol/L LiOH solution at 40°C [13]. In spite of the significance of $Li_2CO_3$ production and growing environmental concerns emphasizing the development of carbon capture reactions, studies on $Li_2CO_3$ production using $CO_{2(g)}$ remain limited.

In recent years, high-throughput experimentation (HTE) has been widely adopted for materials discovery and optimization across various fields [14–18]. HTE accelerates discovery and optimization processes by conducting multiple small-scale experiments simultaneously. However, utilizing HTE without the intelligent selection of parameters and experiments can still be both costly and challenging. The integration of artificial intelligence (AI) tools, particularly active learning, has greatly advanced the field of materials science [19–26]. This integration has given rise to AI-enabled HTE platforms, which offer an effective way to intelligently navigate chemical space using active learning techniques [15,19,27–29]. Gaussian Process Regression (GPR) models are among the most popular machine learning tools for materials process optimization and active learning platforms [30–34]. Their popularity stems from their flexibility, predictive power, and Bayesian nature, which enables effective uncertainty analysis [35–37].

In this study, we propose a Bayesian active learning-driven high-throughput workflow to optimize the $CO_{2(g)}$-based lithium brine softening method for producing solid lithium carbonate, tailored for the battery industry. Using a simplified representation of the system that only included the chemical nature of the compounds, we were able to monitor changes in the system and define traceable optimization parameters. In the following, the design of this workflow and our key findings are discussed in detail.

The active learning cycle, illustrated in **Figure 1**, began with high throughput experiments conducted at 66°C. This temperature was chosen to maximize the $Li_2CO_3$ crystallization rate [7,8,12]. To effectively implement the HTE, it was necessary to extensively monitor system progress, including tracking the concentration of various species. However, considering the complexity of the system of equilibria in this study, this process required significant resources and automation, leading to increased cost and time. To mitigate these challenges, we narrowed our exploration scope to focus on major elements in the system - lithium (Li), nitrogen (N), and carbon (C) - allowing for more efficient and swift monitoring of their concentrations before and after reactions. The details of characterization methods and related measurements are discussed in detail in the supplementary information.

Each HTE batch consisted of 24 miniature experiments, each with an 8 mL volume, and focused on the softening reaction between $NH_4OH$, $CO_{2(g)}$, and LiCl. The experimental procedure, as shown in **Figure 2**, involved bubbling $CO_{2(g)}$ into an ammonium hydroxide solution, followed by mixing it with a LiCl solution in different ratios to initiate crystallization. We conducted initial measurements of C, N, and Li concentrations in the input batches using infrared (IR) spectroscopy and ion chromatography. Subsequent determination of dissolved $Li^+$ ion concentrations in the final solution allowed for calculating sedimented $Li_2CO_3$ and the overall yield in each vial. Further details of the HTE procedures can be found in the supplementary information.

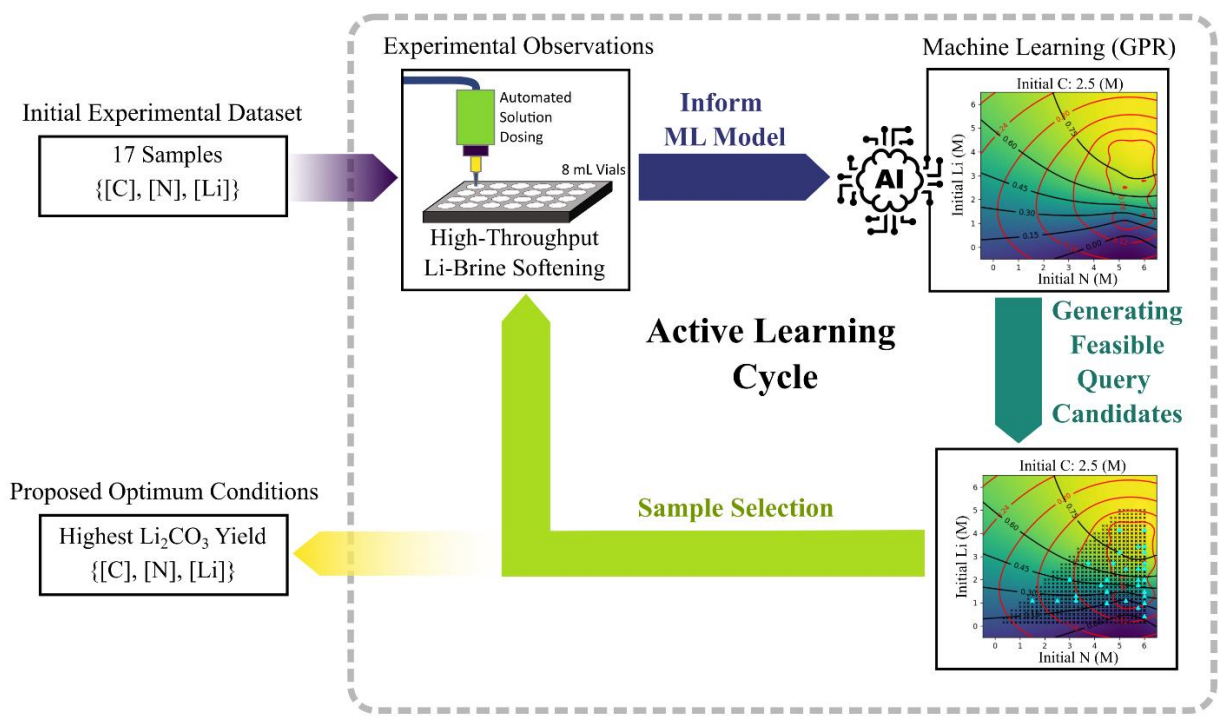

**Figure 1.** The active learning workflow for high-throughput Li-brine softening.

A Gaussian Process Regression (GPR) model was used as the core of the predictive analytics tool. The model, trained using all available experimental data at each iteration, predicts lithium yields within a defined chemical space of nitrogen, carbon, and lithium (N-C-Li). This N-C-Li space, meshed within a range of 0 to 6 mol/L, informed our data acquisition strategy, which segmented the data pool into three tiers: high lithium carbonate yield, large uncertainty in GPR predictions, and random exploration. Supplementary **Figure S2** and **Table S2** detail the experimental limitations and the conditions viable for data acquisition.

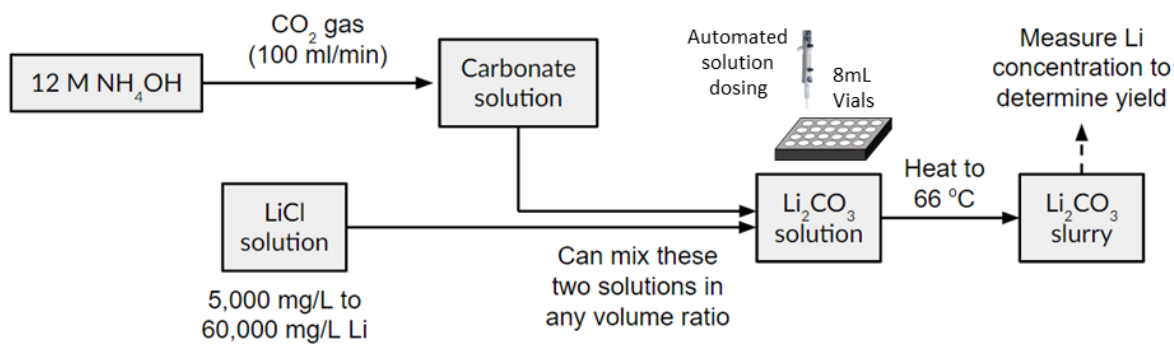

**Figure 2.** Block flow diagram of the high-throughput experimentation workflow and softening process for converting LiCl to $Li_2CO_3$ using $CO_{2(g)}$ and aqueous ammonium hydroxide ($NH_4OH$).

The AI-enhanced HTE approach facilitated a comprehensive exploration of the lithium brine softening process, effectively predicting lithium carbonate yields within the initial N-C-Li parameter space, as depicted in **Figure S3**. Remarkably, just two iterations of AI-driven HTE cycles were sufficient to identify yield values of 83% surpassing the 65% yield reported from traditional lab techniques. This yield is 10% higher than the highest yield reported by Y. Sun et al. for $Li_2CO_3$ production using $CO_{2(g)}$[38]. Notably, the optimized 83% Li2CO3 yield achieved in this study not only matches or exceeds yields obtained using combinations of $CO_2(g)$ with sodium- or potassium-hydroxide but also, does so at 20 to 30 degrees Celsius lower [7,39–41]. This lower temperature approach not only reduces energy consumption but also attains yields that are either higher or comparable, while efficiently removing excess sodium and potassium impurities from the material matrix and further minimizing the carbon footprint.

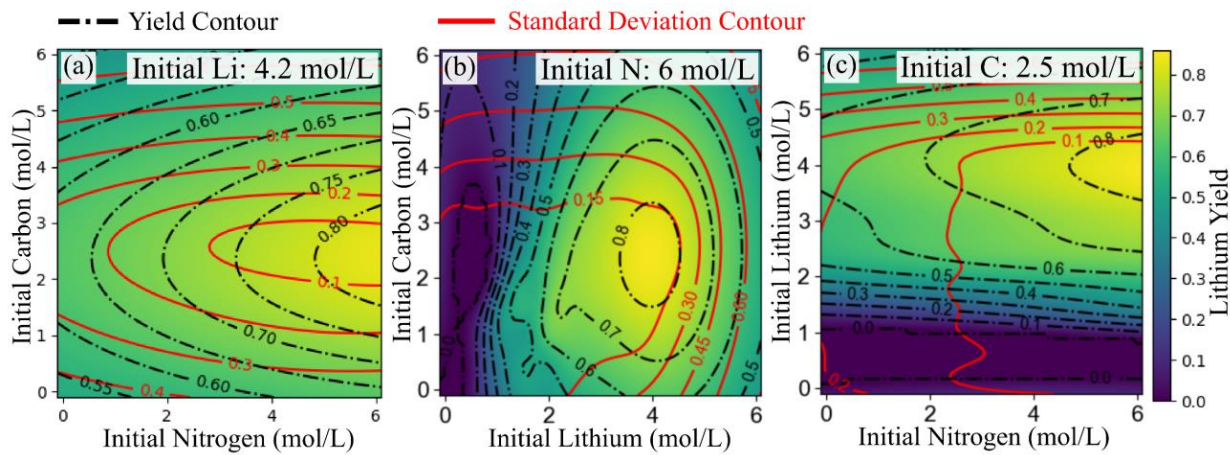

**Figure 3.** Lithium carbonate yield predictions for varying lithium brine softening conditions using the GPR model. This figure showcases the predicted yield across different slices of the chemical landscape, leveraging both historic and newly acquired data from the active learning cycles. Each panel visualizes the yield for a set initial concentration: (a) 4.2 mol/L lithium, (b) 6 mol/L nitrogen, and (c) 2.5 mol/L carbon, with other elements varying from 0 to 6 mol/L. The color gradient indicates the lithium yield, while black contours emphasize differences in yield regions and red contours delineate areas of standard deviation within the GPR predictions, thereby representing the model's uncertainty.

Slices of the explored chemical space (**Figure 3**) highlight conditions for maximizing lithium carbonate yields as predicted by our AI-enhanced HTE approach. Despite the potential for even higher yields suggested by active learning, experimental constraints, such as pipe blockages and sedimentation in the vials due to high lithium carbonate concentrations, limited further exploration. However, these observations hinted at further potential for improvement.

As illustrated in **Figure 3,** our active learning model underscores the significance of high initial lithium concentrations for achieving superior yields. It suggests that optimal lithium concentrations are slightly below the stoichiometric ratio of lithium to carbon in $Li_2CO_3$, ideally at 4.2 molar for initial lithium and approximately 2.5 molar for initial carbon, suggesting an excess of carbon. In addition, Figure 3 shows that achieving high yields requires nitrogen concentrations to significantly exceed stoichiometric expectations for $CO_2$ capture, contrary to the initial chemical intuition of equivalent nitrogen and carbon concentrations. Interestingly, higher initial carbon concentrations reduced yield, defying the expectation of increased yield from higher carbonate concentration as per Le Chatelier's principle. This reduction in yield is likely due to a drop in pH following $CO_2$ dissolution, where each mole of carbon releases two moles of protons during carbonic acid neutralization.

To evaluate the performance of our Gaussian Process Regression models, we monitored the uncertainty of predictions across iterations. The evolving landscape of the predicted yield space is presented in **Figure S4**. A notable decrease in prediction uncertainty after the first active learning cycle is evident in **Figure S5**, underscoring the method's ability to rapidly identify optimal initial reaction conditions in the space of N-C-Li. The second iteration did not show a significant reduction in prediction uncertainty, indicating that this cycle focused more on refining and validating the model's initial predictions. This is corroborated by trends in **Figure S5**, where the area with low prediction uncertainty expands, but the yield landscape remains relatively unchanged.

In summary, this study demonstrates that active learning can accelerate the optimization of chemical reactions occurring in complex systems of equilibria. Reducing the chemical space to key elements has proven instrumental in enabling swift and effective system optimization. By simplifying the chemical space, we were able to define and track reaction parameters, which led to the development of our AI-enhanced HTE workflow. Aligned with the established performance of GPR models reported in materials optimization processes ([20,31–34,37]), our active learning workflow also benefited from this powerful methodology. This innovative method facilitated efficient exploration and optimization of the chemical space, culminating in lithium carbonate yields exceeding 83% in just a single iteration of our AI-boosted HTE cycle. This yield marks a substantial improvement over existing yields reported in the literature for $Li_2CO_3$ production from LiCl brines using $NH_4OH$ and $CO_2$ ([7,38]). Looking ahead, the methodologies and insights from this study hold wide applicability, extending to various reactions and fields within material science. Future endeavors will focus on further refining these AI-driven techniques, applying them to other complex chemical processes, and enhancing their scalability and adaptability to meet diverse industrial requirements.

**ASSOCIATED CONTENT**

Supporting Information is available at the publisher's website.

Experimental procedures, experimental table of experimental results, complementary analyses, executable Python scripts and data can be found at the GitHub repo devoted to this project (https://github.com/shmouses/AI-enabled-HTE-Li-Production).

**AUTHOR INFORMATION**

Corresponding Author:

Jason Hattrick-Simpers – Canmet MATERIALS, Natural Resources Canada, 183 Longwood Rd S, Hamilton, ON, Canada; orcid.org/0000-0003-2937-3188

Email: jason.hattrick-simpers@nrcan-rncan.gc.ca

Authors:

S. Shayan Mousavi Masouleh – Canmet MATERIALS, Natural Resources Canada, 183 Longwood Rd S, Hamilton, ON, Canada; Department of Materials Science and Engineering, McMaster University, 1280 Main St W, Hamilton, ON, Canada; orcid.org/0000-0003-0313-7590

Corey A. Sanz – Telescope Innovations, 301-2386 E Mall, Vancouver, BC, Canada; orcid.org/0000-0002-3836-0744

Ryan P. Jansonius – Telescope Innovations, 301-2386 E Mall, Vancouver, BC, Canada; orcid.org/0000-0002-4014-2068

Samuel Shi – Department of Materials Science and Engineering, University of Toronto, 184 College St, Toronto, ON, Canada;

Maria J. Gendron Romero – Department of Materials Science and Engineering, University of Toronto, 184 College St, Toronto, ON, Canada;

Jason E. Hein – Telescope Innovations, 301-2386 E Mall, Vancouver, BC, Canada; orcid.org/0000-0002-4345-3005

**Notes**

The authors declare no competing financial interest.

**ACKNOWLEDGMENT**

The authors gratefully acknowledge funding from the Critical Minerals Research, Development, and Demonstration (CMRDD) Program administered by Natural Resources Canada. We also extend our gratitude to Standard Lithium for their financial support and valuable discussions that have guided the project's design. Additionally, Telescope Innovations thanks the Mining Innovation Commercialization Accelerator (MICA) for financial support related to this project.

**REFERENCES**

(1)     Meng, F.; McNeice, J.; Zadeh, S. S.; Ghahreman, A. Review of Lithium Production and Recovery from Minerals, Brines, and Lithium-Ion Batteries. *Mineral Processing and Extractive Metallurgy Review* **2021**, *42* (2), 123–141. https://doi.org/10.1080/08827508.2019.1668387.

(2)     Li, L.; Deshmane, V. G.; Paranthaman, M. P.; Bhave, R.; Moyer, B. A.; Harrison, S. Lithium Recovery from Aqueous Resources and Batteries: A Brief Review. *Johnson Matthey Technology Review* **2018**, *62* (2), 161–176.

(3)     Ma, Y.; Svärd, M.; Xiao, X.; Gardner, J. M.; Olsson, R. T.; Forsberg, K. Precipitation and Crystallization Used in the Production of Metal Salts for Li-Ion Battery Materials: A Review. *Metals* **2020**, *10* (12), 1609.

(4)     Yang, W.; Zhou, L.; Dai, J.; Zhou, L.; Zhang, M.; Xie, C.; Hao, H.; Hou, B.; Bao, Y.; Yin, Q. Crystallization of Lithium Carbonate from Aqueous Solution: New Insights into Crystal Agglomeration. *Ind. Eng. Chem. Res.* **2019**, *58* (39), 18448–18455. https://doi.org/10.1021/acs.iecr.9b03644.

(5)     Zhao, S.; Gao, J.; Ma, S.; Li, C.; Ma, Y.; He, Y.; Gong, J.; Zhou, F.; Zhang, B.; Tang, W. Mechanism and Modelling of Reactive Crystallization Process of Lithium Carbonate. *Processes* **2019**, *7* (5), 248.

(6)     Hein, J. E.; Kennepohl, J. P. Process and Method for Refining Lithium Carbonate Starting from an Impure Lithium Chloride Solution, June 17, 2021. https://patents.google.com/patent/US20210180153A1/en (accessed 2024-01-17).

(7)     Kim, S.; Yoon, H.; Min, T.; Han, B.; Lim, S.; Park, J. Carbon Dioxide Utilization in Lithium Carbonate Precipitation: A Short Review. *Environmental Engineering Research* **2024**, *29* (3). https://doi.org/10.4491/eer.2023.553.

(8)     Tian, M.; Wang, Z.; Cao, J.; Guo, J.; Gong, X. Insight into Lithium Carbonate Crystallization in the Mild Reaction System LiCl-NH3· H2O-CO2 by Stabilizing the Solution with NH3· H2O. *Journal of Crystal Growth* **2019**, *520*, 46–55.

(9)     Tanimura, Y.; Mitsuhashi, K.; Kawarabuki, R.; Kawata, M.; Yamaguchi, Y. Method for Producing Lithium Carbonate, December 30, 2014. https://patents.google.com/patent/US8920763B2/en (accessed 2024-01-17).

(10)    Ryabtsev, A. D.; Menzheres, L. T.; Kurakov, A. A.; Gushchina, E. P. Interaction of Ammonium Bicarbonate with Lithium Chloride Solutions. *Theor Found Chem Eng* **2006**, *40* (6), 649–654. https://doi.org/10.1134/S0040579506060157.

(11)    Zevenhoven, R.; Eloneva, S.; Teir, S. Chemical Fixation of CO2 in Carbonates: Routes to Valuable Products and Long-Term Storage. *Catalysis Today* **2006**, *115* (1–4), 73–79.

(12)    Tian, M.; Guo, J.; Wang, Z.; Cao, J.; Gong, X. Synergetic Effect of Secondary Nucleation and Growth on the Lithium Carbonate Particle Size in the Gas–Liquid Reactive Crystallization of LiCl–NH3·H2O–CO2. *Particuology* **2020**, *51*, 10–17. https://doi.org/10.1016/j.partic.2019.10.006.

(13)    Zhou, Z.; Liang, F.; Qin, W.; Fei, W. Coupled Reaction and Solvent Extraction Process to Form Li2CO3: Mechanism and Product Characterization. *AIChE Journal* **2014**, *60* (1), 282–288. https://doi.org/10.1002/aic.14243.

(14)    Malig, T. C.; Yunker, L. P. E.; Steiner, S.; Hein, J. E. Online High-Performance Liquid Chromatography Analysis of Buchwald–Hartwig Aminations from within an Inert Environment. *ACS Catal.* **2020**, *10* (22), 13236–13244. https://doi.org/10.1021/acscatal.0c03530.

(15)    Wang, Y.; Goh, B.; Nelaturu, P.; Duong, T.; Hassan, N.; David, R.; Moorehead, M.; Chaudhuri, S.; Creuziger, A.; Hattrick-Simpers, J.; Thoma, D. J.; Sridharan, K.; Couet, A. Integrated High-Throughput and Machine Learning Methods to Accelerate Discovery of Molten

Salt Corrosion-Resistant Alloys. *Advanced Science* **2022**, *9* (20), 2200370. https://doi.org/10.1002/advs.202200370.

(16)	Green, M. L.; Choi, C. L.; Hattrick-Simpers, J. R.; Joshi, A. M.; Takeuchi, I.; Barron, S. C.; Campo, E.; Chiang, T.; Empedocles, S.; Gregoire, J. M. Fulfilling the Promise of the Materials Genome Initiative with High-Throughput Experimental Methodologies. *Applied Physics Reviews* **2017**, *4* (1).

(17)	DeCost, B.; Joress, H.; Sarker, S.; Mehta, A.; Hattrick-Simpers, J. Towards Automated Design of Corrosion Resistant Alloy Coatings with an Autonomous Scanning Droplet Cell. *JOM* **2022**, *74* (8), 2941–2950.

(18)	Shevlin, M. Practical High-Throughput Experimentation for Chemists. *ACS medicinal chemistry letters* **2017**, *8* (6), 601–607.

(19)	Vasudevan, R. K.; Choudhary, K.; Mehta, A.; Smith, R.; Kusne, G.; Tavazza, F.; Vlcek, L.; Ziatdinov, M.; Kalinin, S. V.; Hattrick-Simpers, J. Materials Science in the Artificial Intelligence Age: High-Throughput Library Generation, Machine Learning, and a Pathway from Correlations to the Underpinning Physics. *MRS communications* **2019**, *9* (3), 821–838.

(20)	DeCost, B. L.; Hattrick-Simpers, J. R.; Trautt, Z.; Kusne, A. G.; Campo, E.; Green, M. L. Scientific AI in Materials Science: A Path to a Sustainable and Scalable Paradigm. *Machine learning: science and technology* **2020**, *1* (3), 033001.

(21)	Li, J.; Lim, K.; Yang, H.; Ren, Z.; Raghavan, S.; Chen, P.-Y.; Buonassisi, T.; Wang, X. AI Applications through the Whole Life Cycle of Material Discovery. *Matter* **2020**, *3* (2), 393–432.

(22)     López, C. Artificial Intelligence and Advanced Materials. *Advanced Materials* **2023**, *35* (23), 2208683. https://doi.org/10.1002/adma.202208683.

(23)     Sun, S.; Hartono, N. T. P.; Ren, Z. D.; Oviedo, F.; Buscemi, A. M.; Layurova, M.; Chen, D. X.; Ogunfunmi, T.; Thapa, J.; Ramasamy, S.; Settens, C.; DeCost, B. L.; Kusne, A. G.; Liu, Z.; Tian, S. I. P.; Peters, I. M.; Correa-Baena, J.-P.; Buonassisi, T. Accelerating Photovoltaic Materials Development via High-Throughput Experiments and Machine-Learning-Assisted Diagnosis. *Joule* **2019**, *3* (6), 1437–1451. https://doi.org/10.1016/j.joule.2019.05.014.

(24)     Back, S.; Aspuru-Guzik, A.; Ceriotti, M.; Gryn'ova, G.; Grzybowski, B.; Ho Gu, G.; Hein, J.; Hippalgaonkar, K.; Hormázabal, R.; Jung, Y.; Kim, S.; Youn Kim, W.; Mohamad Moosavi, S.; Noh, J.; Park, C.; Schrier, J.; Schwaller, P.; Tsuda, K.; Vegge, T.; Lilienfeld, O. A. von; Walsh, A. Accelerated Chemical Science with AI. *Digital Discovery* **2024**, *3* (1), 23–33. https://doi.org/10.1039/D3DD00213F.

(25)     Mousavi M, S. S.; Pofelski, A.; Botton, G. EELSpecNet: Deep Convolutional Neural Network Solution for Electron Energy Loss Spectroscopy Deconvolution. *Microscopy and Microanalysis* **2021**, *27* (S1), 1626–1627. https://doi.org/10.1017/S1431927621005997.

(26)     Mousavi M, S. S.; Pofelski, A.; Teimoori, H.; Botton, G. A. Alignment-Invariant Signal Reality Reconstruction in Hyperspectral Imaging Using a Deep Convolutional Neural Network Architecture. *Scientific Reports* **2022**, *12* (1), 17462.

(27)     Ren, F.; Ward, L.; Williams, T.; Laws, K. J.; Wolverton, C.; Hattrick-Simpers, J.; Mehta, A. Accelerated Discovery of Metallic Glasses through Iteration of Machine Learning and High-Throughput Experiments. *Science Advances* **2018**, *4* (4), eaaq1566. https://doi.org/10.1126/sciadv.aaq1566.

(28)	Bunn, J. K.; Han, S.; Zhang, Y.; Tong, Y.; Hu, J.; Hattrick-Simpers, J. R. Generalized Machine Learning Technique for Automatic Phase Attribution in Time Variant High-Throughput Experimental Studies. *Journal of Materials Research* **2015**, *30* (7), 879–889. https://doi.org/10.1557/jmr.2015.80.

(29)	Wang, A.; Liang, H.; McDannald, A.; Takeuchi, I.; Kusne, A. G. Benchmarking Active Learning Strategies for Materials Optimization and Discovery. *Oxford Open Materials Science* **2022**, *2* (1), itac006.

(30)	Williams, C. K.; Rasmussen, C. E. *Gaussian Processes for Machine Learning*; MIT press Cambridge, MA, 2006; Vol. 2.

(31)	Chen, J.; Kang, L.; Lin, G. Gaussian Process Assisted Active Learning of Physical Laws. *Technometrics* **2021**, *63* (3), 329–342. https://doi.org/10.1080/00401706.2020.1817790.

(32)	Noack, M. M.; Doerk, G. S.; Li, R.; Streit, J. K.; Vaia, R. A.; Yager, K. G.; Fukuto, M. Autonomous Materials Discovery Driven by Gaussian Process Regression with Inhomogeneous Measurement Noise and Anisotropic Kernels. *Sci Rep* **2020**, *10* (1), 17663. https://doi.org/10.1038/s41598-020-74394-1.

(33)	Uteva, E.; Graham, R. S.; Wilkinson, R. D.; Wheatley, R. J. Active Learning in Gaussian Process Interpolation of Potential Energy Surfaces. *The Journal of Chemical Physics* **2018**, *149* (17), 174114. https://doi.org/10.1063/1.5051772.

(34)	Deringer, V. L.; Bartók, A. P.; Bernstein, N.; Wilkins, D. M.; Ceriotti, M.; Csányi, G. Gaussian Process Regression for Materials and Molecules. *Chem. Rev.* **2021**, *121* (16), 10073–10141. https://doi.org/10.1021/acs.chemrev.1c00022.

(35)   Gongora, A. E.; Xu, B.; Perry, W.; Okoye, C.; Riley, P.; Reyes, K. G.; Morgan, E. F.; Brown, K. A. A Bayesian Experimental Autonomous Researcher for Mechanical Design. *Sci. Adv.* **2020**, *6* (15), eaaz1708. https://doi.org/10.1126/sciadv.aaz1708.

(36)   Biswas, A.; Liu, Y.; Creange, N.; Liu, Y.-C.; Jesse, S.; Yang, J.-C.; Kalinin, S. V.; Ziatdinov, M. A.; Vasudevan, R. K. A Dynamic Bayesian Optimized Active Recommender System for Curiosity-Driven Human-in-the-Loop Automated Experiments. arXiv April 5, 2023. http://arxiv.org/abs/2304.02484 (accessed 2024-01-28).

(37)   Ziatdinov, M.; Liu, Y.; Kelley, K.; Vasudevan, R.; Kalinin, S. V. Bayesian Active Learning for Scanning Probe Microscopy: From Gaussian Processes to Hypothesis Learning. *ACS Nano* **2022**, *16* (9), 13492–13512. https://doi.org/10.1021/acsnano.2c05303.

(38)   Sun, Y.; Song, X.; Wang, J.; Yu, J. Preparation of $Li_2CO_3$ by Gas-liquid Reactive Crystallization of LiOH and $CO_2$. *Cryst. Res. Technol.* **2012**, *47* (4), 437–442. https://doi.org/10.1002/crat.201100571.

(39)   Battaglia, G.; Berkemeyer, L.; Cipollina, A.; Cortina, J. L.; Fernandez De Labastida, M.; Lopez Rodriguez, J.; Winter, D. Recovery of Lithium Carbonate from Dilute Li-Rich Brine via Homogenous and Heterogeneous Precipitation. *Ind. Eng. Chem. Res.* **2022**, *61* (36), 13589–13602. https://doi.org/10.1021/acs.iecr.2c01397.

(40)   Han, B.; Haq, R. A. U.; Louhi-Kultanen, M. Lithium Carbonate Precipitation by Homogeneous and Heterogeneous Reactive Crystallization. *Hydrometallurgy* **2020**, *195*, 105386.

(41)   Jiang, Y.; Liu, C.; Zhou, X.; Li, P.; Song, X.; Yu, J. Toward CO2 Utilization: Gas–Liquid Reactive Crystallization of Lithium Carbonate in Concentrated KOH Solution. *Energy Sources,*

*Part A: Recovery, Utilization, and Environmental Effects* **2021**, *43* (24), 3332–3344. https://doi.org/10.1080/15567036.2019.1587068.

# Artificial Intelligence-Enabled Optimization of Battery-Grade Lithium Carbonate Production

S. Shayan Mousavi Masouleh [1,2], Corey A. Sanz [3], Ryan P. Jansonius [3], Samuel Shi [4], Maria J. Gendron Romero [4], Jason E. Hein [3], Jason Hattrick-Simpers [1,*]

[1] Canmet MATERIALS, Natural Resources Canada, 183 Longwood Rd S, Hamilton, ON, Canada

[2] Department of Materials Science and Engineering, McMaster University, 1280 Main St W, Hamilton, ON, Canada

[3] Telescope Innovations, 301-2386 E Mall, Vancouver, BC, Canada

[4] Department of Materials Science and Engineering, University of Toronto, 184 College St, Toronto, ON, Canada

**Corresponding Author**

* jason.hattrick-simpers@nrcan-rncan.gc.ca

# Supplementary Information:

## 1. Experimental procedures:

### 1.1. Materials

All chemicals were purchased from Sigma Aldrich (purity >98%). DI water ( >13 MOhm) was sourced from an in-house reverse osmosis unit. CO2 gas was obtained from Praxair.

### 1.2. Instrumentation

Ion chromatography (IC) data was recorded on a Dionex ICS-6000 equipped with a CS16-4uM, 4X250mm column. Eluent consisted of isocratic 35 mM methane sulfonic acid with a flow rate of 0.640 ml/min. Liquid dosing for high throughput experiments was carried out with an Unchained Labs Junior pipette robot. Samples were mixed and heated with an IKA ThermoShaker orbital mixer. Infrared (IR) spectroscopy measurements were made using an MT ReactIR probe.

### 1.3. Procedure for 8 ml vial experiments:

The experiments were carried out in 8 ml vials in a 24 well plate and the data that was collected is tabled in Table S1. Solutions were made by mixing aqueous LiCl with a solution of $NH_4OH$ that has been carbonated by bubbling $CO_2$ into it (see below for details on this solution). Aliquots from the $NH_4OH/CO_2$ solution were removed at various time intervals of elapsed $CO_2$ bubbling time in order to collect data at different carbonation levels (measured as total carbon concentration).

Another variable that was altered was the ratio between the volume of LiCl solution added and the volume of the $NH_4OH/CO_2$ solution added (this effectively varies the total nitrogen concentration in each sample). The final variable that was altered was the Li concentration in the LiCl solution. All three of these variables were varied in each sample to give rise to solutions that contained different concentrations of Li, C, and N (see Figure 3).

The vials were then capped and heated to 66 °C, while mixing vigorously in an orbital mixer. Note that heating is required to speed up the crystallization - even though $Li_2CO_3$ solubility does not change with temperature in this system, the vials are still heated to complete the experiment within 24 hours. After mixing at 66 °C for 24 hours, a 1 ml aliquot of the solution from each sample (often a slurry of $Li_2CO_3$) was removed and filtered through a syringe filter (note that cooling during filtration is not an issue for this system because the solubility of $Li_2CO_3$ doesn't change with temperature). The filtered solution was then diluted, and the Li concentration was measured by ion chromatography.

**Table S1.** This table presents data from high-throughput experiments utilized to develop the yield optimization model. The first column lists the batch number from which the data were obtained. Batch 0 contains historical data acquired through the traditional grid search method. Data from batches 1 and 2 were generated during active learning cycles. The acquisition policy that suggested each input combination is noted in the rightmost column. Total concentrations of C, N, and Li were calculated based on the concentration of these elements in the solution mix before crystallization occurred. This involves proper dilution calculations from the initial unmixed LiCl and $NH_4OH/CO_2$ solutions. Total C concentration was adjusted by sampling from the carbonation reactor at various times after initiating $CO_2$ bubbling. Total N and Li concentrations were altered by mixing the $NH_4OH/CO_2$ solution with the LiCl solution in different volume ratios, or by changing the initial Li concentration in the LiCl solution. Data in the table are rounded to two digits after the decimal for clarity.

| # Batch | Initial aqueous concentrations (before crystallization, M) | | | Measured Li concentration at equilibrium (M) | %Yield | Acquisition Policy |
|---|---|---|---|---|---|---|
| | Total Carbon | Total Nitrogen | Total Lithium | | | |
| 0 | 0.50 | 4.50 | 1.00 | 0.72 | 0.28 | Grid Search |
| 0 | 1.00 | 4.50 | 1.00 | 0.66 | 0.34 | Grid Search |
| 0 | 1.50 | 4.50 | 1.00 | 0.74 | 0.26 | Grid Search |
| 0 | 1.00 | 6.00 | 1.00 | 0.62 | 0.38 | Grid Search |
| 0 | 1.50 | 6.00 | 1.00 | 0.66 | 0.34 | Grid Search |
| 0 | 0.50 | 4.50 | 1.50 | 0.74 | 0.50 | Grid Search |
| 0 | 1.00 | 4.50 | 1.50 | 0.68 | 0.55 | Grid Search |
| 0 | 1.50 | 4.50 | 1.50 | 0.74 | 0.51 | Grid Search |
| 0 | 0.50 | 6.00 | 1.50 | 0.75 | 0.50 | Grid Search |
| 0 | 1.00 | 6.00 | 1.50 | 0.82 | 0.46 | Grid Search |
| 0 | 1.50 | 6.00 | 1.50 | 0.68 | 0.55 | Grid Search |
| 0 | 0.50 | 4.50 | 2.00 | 1.01 | 0.50 | Grid Search |
| 0 | 1.00 | 4.50 | 2.00 | 1.04 | 0.48 | Grid Search |
| 0 | 1.50 | 4.50 | 2.00 | 0.78 | 0.61 | Grid Search |
| 0 | 0.50 | 6.00 | 2.00 | 0.98 | 0.51 | Grid Search |
| 0 | 1.00 | 6.00 | 2.00 | 0.72 | 0.64 | Grid Search |
| 0 | 1.50 | 6.00 | 2.00 | 0.69 | 0.65 | Grid Search |
| 1 | 2.49 | 6.00 | 2.74 | 0.73 | 0.73 | High Yield |
| 1 | 1.98 | 6.00 | 2.50 | 0.71 | 0.72 | High Yield |
| 1 | 2.54 | 5.25 | 2.50 | 0.76 | 0.70 | High Yield |
| 1 | 2.17 | 5.75 | 2.02 | 0.78 | 0.61 | High Yield |
| 1 | 2.55 | 5.75 | 3.45 | 0.69 | 0.80 | High Yield |
| 1 | 2.12 | 6.00 | 3.45 | 0.67 | 0.81 | High Yield |
| 1 | 2.56 | 5.00 | 3.21 | 0.81 | 0.75 | High Yield |
| 1 | 1.61 | 6.00 | 3.21 | 0.78 | 0.76 | High Yield |
| 1 | 1.71 | 5.75 | 1.79 | 0.82 | 0.54 | High Yield |
| 1 | 1.36 | 5.75 | 2.50 | 0.78 | 0.69 | High Yield |
| 1 | 2.49 | 6.00 | 4.16 | 0.69 | 0.83 | High Yield |
| 1 | 1.68 | 4.75 | 2.74 | 0.79 | 0.71 | High Yield |
| 1 | 2.05 | 4.25 | 1.79 | 0.89 | 0.50 | High Uncertainty |
| 1 | 2.56 | 5.00 | 4.16 | 0.81 | 0.81 | High Uncertainty |
| 1 | 1.81 | 3.75 | 2.74 | 0.85 | 0.69 | High Uncertainty |
| 1 | 0.69 | 3.25 | 1.31 | 0.75 | 0.43 | High Uncertainty |
| 1 | 2.49 | 6.00 | 1.31 | 0.88 | 0.33 | High Uncertainty |
| 1 | 1.90 | 5.75 | 0.80 | 0.80 | 0.00 | High Uncertainty |
| 1 | 0.77 | 1.50 | 1.11 | 0.86 | 0.23 | Random |
| 1 | 1.24 | 3.00 | 2.02 | 0.88 | 0.56 | Random |
| 1 | 2.54 | 5.25 | 1.11 | 0.96 | 0.14 | Random |
| 1 | 1.57 | 3.25 | 1.11 | 0.82 | 0.26 | Random |
| 1 | 0.82 | 2.50 | 1.11 | 0.75 | 0.32 | Random |
| 1 | 0.37 | 6.00 | 0.43 | 0.43 | 0.00 | Random |
| 2 | 0.31 | 5.33 | 0.83 | 0.65 | 0.22 | Random |
| 2 | 0.92 | 2.17 | 2.00 | 0.90 | 0.55 | Random |
| 2 | 0.75 | 3.50 | 0.50 | 0.50 | 0.00 | Random |
| 2 | 1.00 | 3.83 | 2.33 | 0.85 | 0.64 | High Yield |
| 2 | 1.37 | 3.00 | 3.00 | 1.08 | 0.64 | High Yield |
| 2 | 1.64 | 3.83 | 3.50 | 0.98 | 0.72 | High Yield |
| 2 | 1.53 | 5.33 | 3.67 | 0.82 | 0.78 | High Yield |
| 2 | 1.92 | 4.50 | 3.67 | 0.83 | 0.77 | High Yield |
| 2 | 1.99 | 5.00 | 1.83 | 0.84 | 0.54 | High Yield |
| *2 | 0.17 | 0.33 | 0.17 | 0.17 | 0.00 | High Uncertainty |
| *2 | 0.17 | 1.17 | 0.17 | 0.17 | 0.00 | High Uncertainty |
| *2 | 0.17 | 2.00 | 0.17 | 0.17 | 0.00 | High Uncertainty |
| *2 | 0.17 | 2.83 | 0.17 | 0.17 | 0.00 | High Uncertainty |
| *2 | 1.00 | 2.00 | 0.17 | 0.17 | 0.00 | High Uncertainty |
| *2 | 0.17 | 3.67 | 0.17 | 0.17 | 0.00 | High Uncertainty |
| *2 | 2.50 | 6.00 | 0.17 | 0.17 | 0.00 | High Uncertainty |
| *2 | 1.33 | 2.83 | 0.17 | 0.17 | 0.00 | High Uncertainty |
| *2 | 0.00 | 0.00 | 0.00 | 0.00 | 0.00 | High Uncertainty |
| *2 | 2.50 | 5.00 | 0.17 | 0.17 | 0.00 | High Uncertainty |
| *2 | 2.00 | 4.00 | 0.17 | 0.17 | 0.00 | High Uncertainty |
| *2 | 1.17 | 4.83 | 0.33 | 0.33 | 0.00 | Random |
| *2 | 1.00 | 5.67 | 0.17 | 0.17 | 0.00 | Random |
| *2 | 0.33 | 4.50 | 0.17 | 0.17 | 0.00 | Random |

*\* The experiments with batch numbers \*2 were suggested for experimentation by the ML model. However, due to the low concentration of initial lithium and the control experiments conducted by the experimentalist in this study, they were considered extremely diluted. As a result, the output was automatically assumed to be zero.*

### 1.4. Carbonation of 12M ammonium hydroxide solution via $CO_2$ bubbling and measurement of total carbon concentration:

A 50 ml solution of 12M $NH_4OH$ (taken directly from the stock bottle provided by Sigma) was added to a stirred reactor (Rushton turbine impeller used to promote $CO_2$ dissolution). $CO_2$ was bubbled into this solution using a Tygon tube at a rate of 100 ml/min. The temperature of the solution was held at 25 °C and the stir rate was 500 rpm. 100 uL aliquots were removed from the reactor every 15 minutes and diluted into a tube containing 2.5 ml of 1M NaOH and 7.4 ml DI water (10 ml total, 100x dilution) - this is so that all of the dissolved carbon is converted fully over to carbonate so that it can be measured by the IR probe. The total C concentration in each 10 ml diluted sample was measured by ReactIR (specifically the carbonate band at 1395 cm$^{-1}$) using a calibration curve generated using $K_2CO_3$. The carbonate concentration in the diluted sample was multiplied by 100 to determine the total C concentration in the $NH_4OH/CO_2$ solution at the time of aliquot sampling. The data is presented graphically in Figure S1.

For high throughput experiments, the above procedure was repeated, except aliquots were instead removed at various other time intervals (other than every 15 minutes) based on what the predicted total carbon concentration would be at that time. The total C concentration in these aliquots was extrapolated from the data shown in Figure S1.

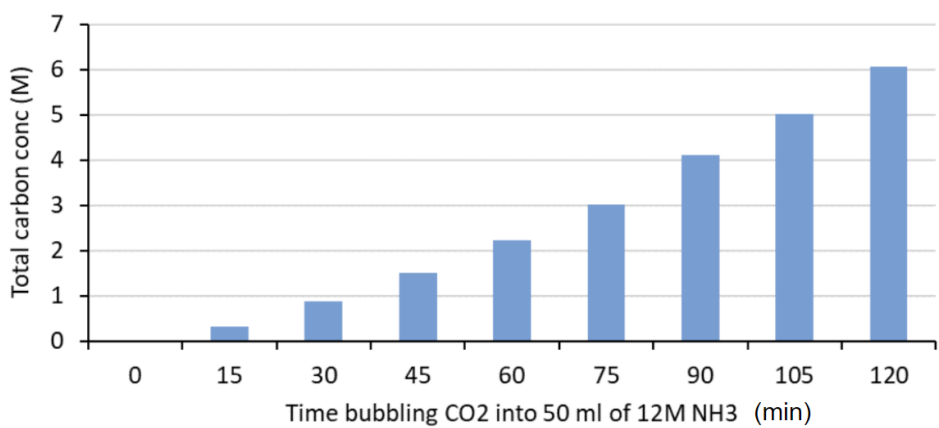

**Figure S1:** Total carbon concentration as determined by IR spectroscopy. Measured every 15 minutes by bubbling $CO_2$ into a 50 ml solution of 12M $NH_4OH$ at a flow rate of 100 ml/min (temperature of reactor = 25 ºC).

## 2. Experimentally viable ranges:

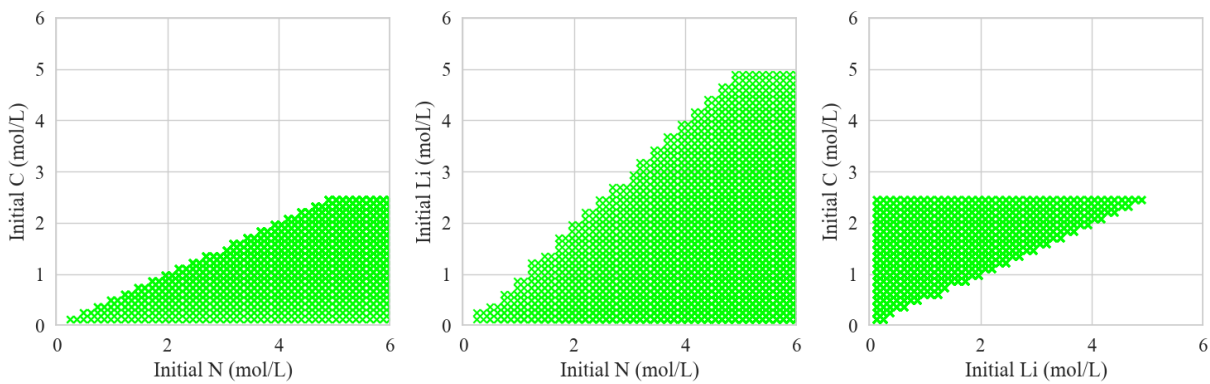

**Figure S2:** The green hashed area identifies the experimentally viable regions for initial concentration and permissible areas for data acquisition. Taking into account specific experimental constraints, such as stoichiometric ratios within the solution and varying solubility levels of different elements in water, not all initial concentrations within the explored chemical space were deemed practical. Further details governing the selection of combinations are provided in Table S2.

**Table S2:** Summary of the rules defining experimentally viable initial concentration combinations for nitrogen, carbon, and lithium in the chemical space. The constraints are categorized into solubility limitations and stoichiometry-controlled relationships, reflecting both the physical properties of the elements and the chemical interactions between them.

| Rule Number | Element(s) | Constraint | Constraint Type |
|---|---|---|---|
| 1 | Nitrogen | Less than 6 mol/L | Solubility Limitation |
| 2 | Carbon | Less than 2.5 mol/L | Solubility Limitation |
| 3 | Lithium | Less than 6 mol/L | Solubility Limitation |
| 4 | Carbon, Nitrogen | Carbon $\leq 0.5 \times$ Nitrogen | Stoichiometry Controlled |
| 5 | Carbon, Lithium | Carbon $\geq 0.5 \times$ Lithium | Stoichiometry Controlled |

Our data acquisition strategy segregates the data pool into three tiers, allocated proportionally at 50%, 25%, and 25%. Each tier samples data based on specific objectives, ensuring that no two sampled points in the chemical space are closer than a Euclidean distance of 1 mol/L. The three acquisition objectives or tiers considered are as follows: (1) lithium carbonate yield of the HTE experiments, where the aim is to acquire initial conditions with the highest predicted lithium yield; (2) uncertainty of the GPR predictions, where the acquisition policy focuses on points with the highest prediction uncertainty; and (3) random exploration of the parameter space, involving random selection of initial condition combinations.

Therefore, beginning with an available initial batch of experimental data (Batch 0), we constructed a preliminary GPR prediction of the chemical space. We conducted two rounds of active learning, with each round proposing 24 experiment suggestions. The initial round (Batch 1) primarily targeted high lithium yield areas, with the majority of experiments sourced from high-yield regions. In contrast, the second round (Batch 2) emphasized broader space exploration based on the GPR model's uncertainty of the predictions (standard deviation). The specific acquisition breakdown for each round is detailed in Table S3.

**Table S3:** Distribution of each data batch according to different acquisition policies.

|         | Acquisition Policy |             |        | Total |
|---------|--------------------|-------------|--------|-------|
|         | Yield              | Uncertainty | Random |       |
| Batch 0 |                    |             |        | 17    |
| Batch 1 | 12                 | 6           | 6      | 24    |
| Batch 2 | 6                  | 12          | 6      | 24    |

## 3. AI-Enhanced Yield Optimization in Li-Brine Softening

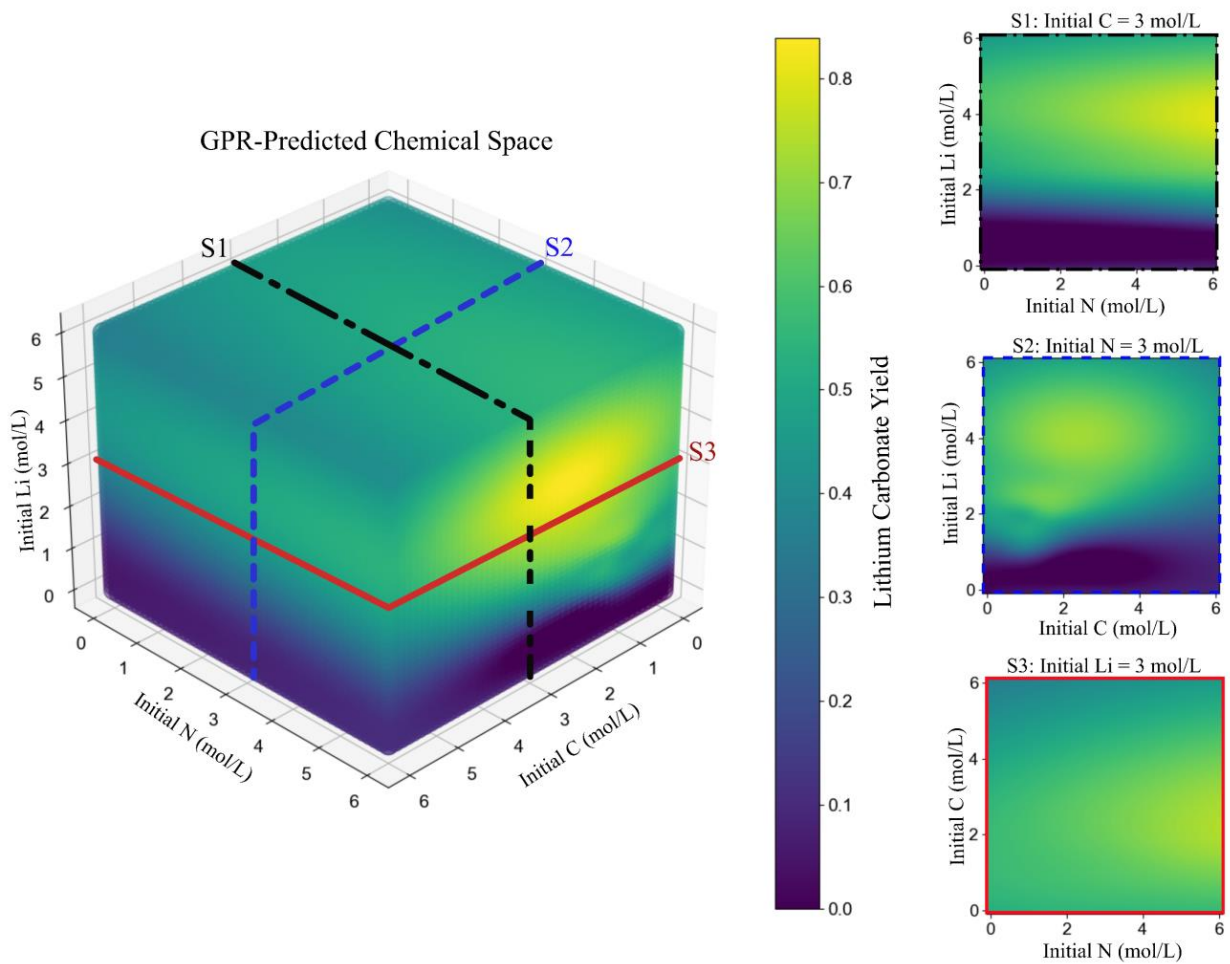

**Figure S3:** Three-dimensional visualization of the predicted chemical space using the GPR model, trained on all collected data (Batches 0, 1, and 2). The space is intentionally under-sampled to highlight the meshed nature of the chemical space. Three specific slices (S1, S2, and S3) are extracted and displayed on the right, providing detailed views of the predicted lithium carbonate yield landscape at different planes within the space.

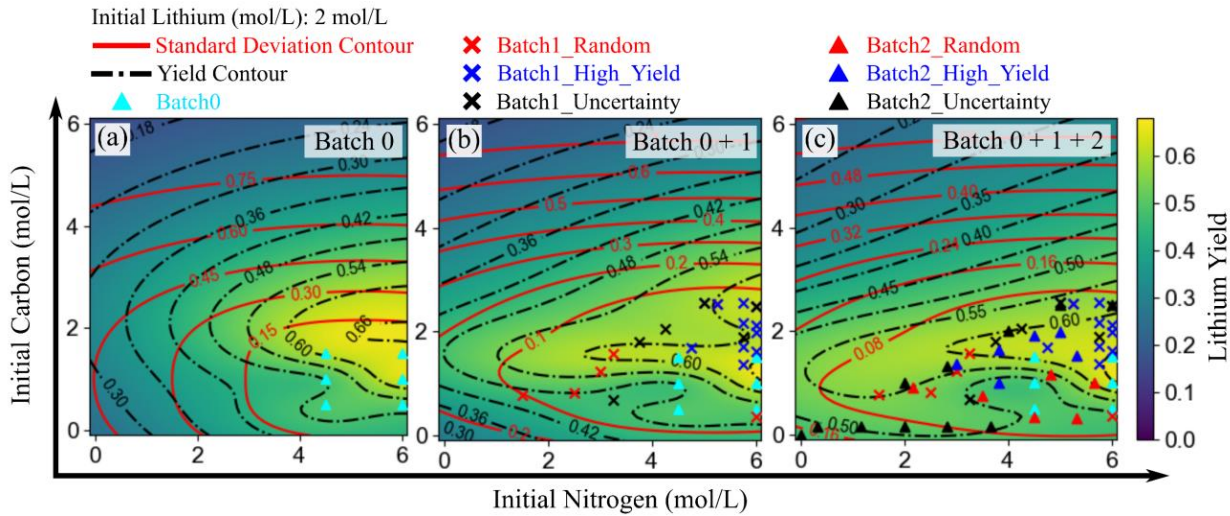

**Figure S4:** Visualization of Lithium Yield Predictions Using Gaussian Process Regression (GPR) in the Context of the Lithium Softening Process. The model was trained on data from three stages: (a) Batch 0 alone, (b) a combination of Batches 0 and 1, and (c) an aggregate of Batches 0, 1, and 2. These stages correspond to iterations 0 to 2 in the active learning cycle. The data points acquired at each stage are overlaid on their respective colormaps. Each colormap captures the lithium yield within a specific chemical space, holding the initial lithium concentration constant at 2 mol/L while allowing for variations in carbon and nitrogen levels between 0 to 6 mol/L. Black and red contours signify lithium yield and the associated GPR prediction uncertainty (measured as standard deviation), respectively.

## 4. Performance Analysis of Active Learning Strategies

During our active learning iterations, we gauged the learning quality by observing the uncertainty in the GPR model's predictions. As shown in Figure 3, uncertainty, measured by the standard deviation of the predictions, reduces with each iteration. Consequently, the areas with low uncertainty (or high prediction certainty) broaden. Using violin plots, we traced the prediction uncertainty throughout each active learning iteration. Additionally, we compared the uncertainty distribution from our tiered data acquisition with that from a fully random acquisition strategy, as illustrated in Figures S6.

The datasets for fully random acquisition, encompassing both initial parameters and predictions, were created using the surrogate GPR model trained on all the available experimental data. To mimic real-world experimental conditions, we introduced a random fluctuation of up to 10% into the model's input parameters, specifically the initial concentrations, as well as the predicted yields. The fully random acquisition distributions shown in Figures S6, represent the average standard deviations of predictions taken over 100 different random seeds for each iteration. Breaking it down further, for iteration 1, we amalgamated the data from batch 0 (consisting of 17 experiments) with data randomly sampled to match the size of batch 1 (24 experiments). This process was repeated across 100 distinct random seeds. Similarly, for the subsequent iteration, data from batch 0 was combined with data randomly drawn to cumulatively match the sizes of batches 1 and 2, amounting to 48 experiments in total. This amalgamation was also repeated across 100 different random seeds to guarantee the representativeness of the acquired datasets.

Figure S6 highlights that the active learning training process effectively reduced the GPR model's uncertainty. This reduction in prediction uncertainty, coupled with the concentration of uncertainties around the mean in the violin plot, suggests an enhanced robustness of the model. Notably, even as the second iteration targeted high-uncertainty areas from the initial predictions, the model remained consistent when exposed to new data. This consistent performance is indicative of the robustness and predictive fidelity of our active learning methodology.

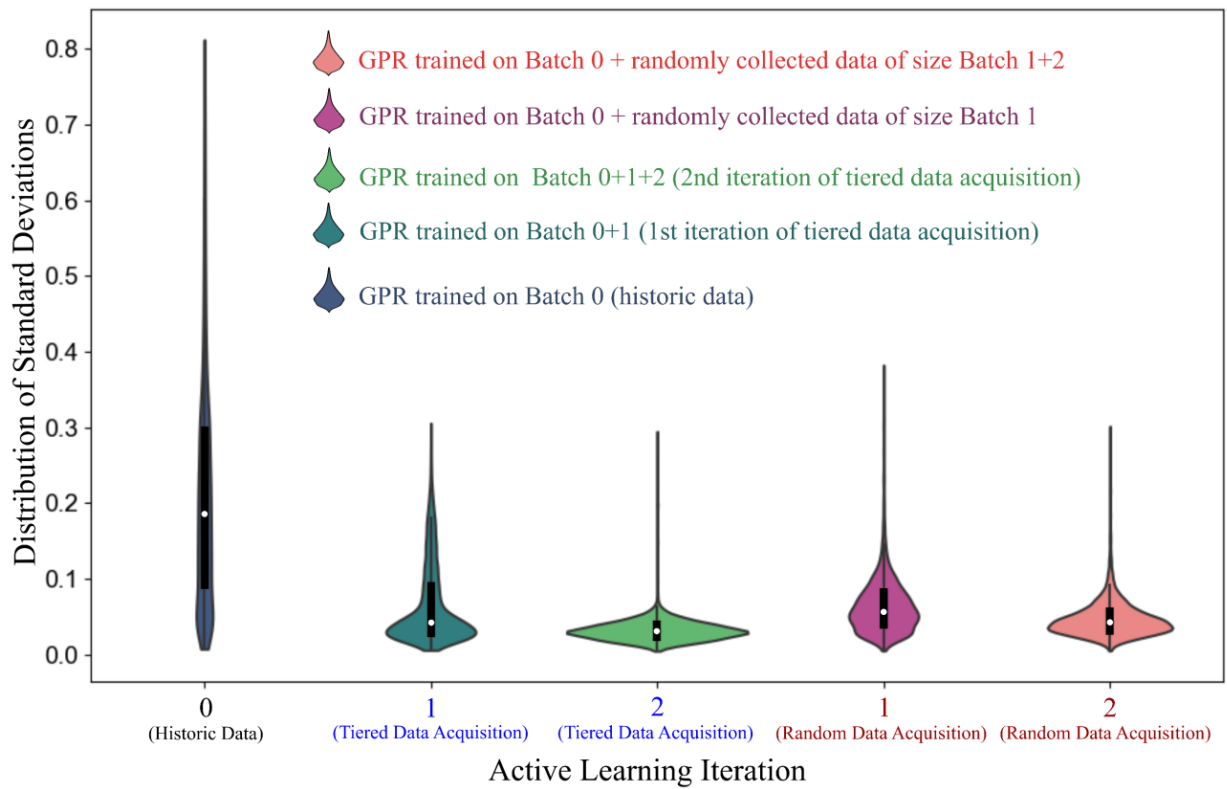

**Figure S5:** Violin plots depicting the overall standard deviation in Gaussian Process Regression (GPR) models across various active learning iterations. The plots compare the performance of models trained using the tiered data acquisition strategy employed in this study with those using fully random data acquisition.

## 5. Evaluation of Data Acquisition Policies

To gauge the influence of each data acquisition tier (high yield, high uncertainty, random) used in this study, we evaluated the individual contributions of each tier. To do this, we trained models using the base data (Batch 0) combined with data sourced through each specific policy per iteration. We used two metrics to quantify the contribution of each acquisition method: mean square error (MSE) and information gain, as depicted in Figure S7.

The MSE metric contrasts a surrogate GPR model—trained using all available data from Batches 0, 1, and 2—against models trained using the base data (Batch 0) combined with data from each specific data acquisition tier (policy). Figure S7 shows that focusing on high-yield areas substantially reduces the MSE of the model, enhancing the accuracy of high-value predictions. Moreover, targeting high uncertainty areas considerably lowers the MSE by addressing regions where the model's predictive certainty is low. In contrast, random data acquisition has a less pronounced effect on model performance when compared to the other two data tiers.

Information gain, a measure derived from information theory, determines the dataset's entropy in relation to a benchmark, which in our case is the entropy of predictions using only the base data (Batch 0). We computed entropy with a continuous differential entropy formula, taking the standard deviation of the GPR predictions to represent local entropy. As displayed in Figure S7, both high yield and uncertainty-focused acquisition strategies provide similar information gains.

It's worth highlighting that although random data acquisition might not have a substantial impact compared to the other methods, excluding it could lead to biased data collection. Even without observing unexpected outcomes in our predictions and chemical space experiments, random data sampling could potentially identify regions prone to such anomalies. Hence, we deem all three data acquisition tiers crucial for this study.

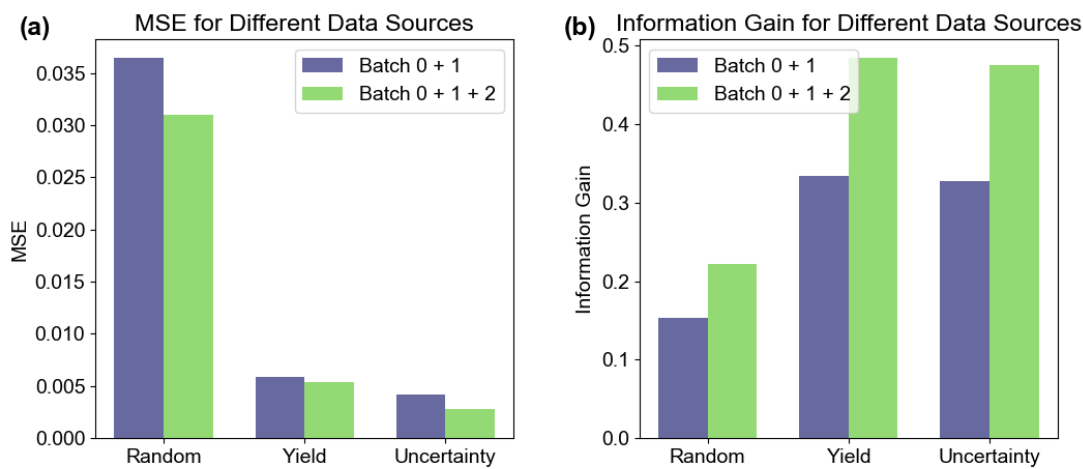

**Figure S6:** Assessment of the performance of various data sources in the active learning process (random acquisition, high yield acquisition, high uncertainty acquisition) using: (a) the mean squared error comparison of models trained with different data sources (acquisition policies) against those trained with all available data; (b) information gain for different data sources (acquisition policies) versus the model trained using only initial data (Batch 0).